\documentclass[]{pasj02} 
\usepackage{natbib} 

\usepackage{xcolor}

\usepackage{etoolbox}
\makeatletter
\pretocmd{\NAT@citex}{\begingroup\color{blue}}{}{}
\apptocmd{\NAT@citex}{\endgroup}{}{}

\pretocmd{\NAT@citexnum}{\begingroup\color{blue}}{}{}
\apptocmd{\NAT@citexnum}{\endgroup}{}{}
\makeatother

\jyear{2026}
\Received{\today}
\Accepted{}

\begin{document} 

\title{Neural Surrogates for Tabulated Nuclear Equations of State in Core-Collapse Supernova Simulations}

\author{
 Youhei \textsc{Masada},\altaffilmark{1}\altemailmark\orcid{0000-0002-0767-3449} \email{ymasada@fukuoka-u.ac.jp} 
 Jin \textsc{Matsumoto},\altaffilmark{1}$^{,\dag}$\orcid{0000-0002-5637-0372}
 and 
 Tomoya \textsc{Takiwaki},\altaffilmark{2}\altemailmark \orcid{0000-0003-0304-9283}
}

\altaffiltext{1}{Department of Applied Physics, Faculty of Science, Fukuoka University, Fukuoka 814-0180, Japan}
\altaffiltext{2}{National Astronomical Observatory of Japan, Osawa, Mitaka, Tokyo 181-8588, Japan}



\KeyWords{High Energy astrophysics -- supernovae –- hydrodynamics -- equation of state -- neural networks}  

\maketitle

\begin{abstract}
Large-scale core-collapse supernova simulations rely on tabulated nuclear equations of state (EOSs), whose memory requirements can become a practical constraint in massively parallel calculations. We explore a compact alternative by replacing the tabulated thermodynamic mapping with a neural-network surrogate, using the Lattimer--Swesty EOS as a representative reference model. The surrogate directly predicts pressure, sound speed, and effective adiabatic index from $(\log \rho, \log e, Y_e)$ and accurately reconstructs the nonlinear thermodynamic structure of the EOS, including its behavior across phase-transition regions. It reproduces one-dimensional shock-tube solutions with hydrodynamic errors at the $\sim 10^{-3}$ level under different thermodynamic conditions and yields hydrostatic equilibrium configurations closely matching those obtained with the reference EOS. We find that smaller regression errors do not necessarily yield better hydrodynamic performance, demonstrating the importance of hydrodynamic validation in surrogate model selection. A compact network with only $\sim 10^4$ parameters retains this accuracy while reducing the memory required to store the EOS from $\sim 160$ MB to $\sim 100$ KB, corresponding to a reduction by a factor of $\sim 1500$. The surrogate also remains stable through collapse, bounce, shock formation, and post-bounce evolution in a one-dimensional adiabatic core-collapse calculation. These results demonstrate the potential of compact neural EOS surrogates for reducing memory requirements in large-scale core-collapse supernova simulations.
\end{abstract}


\section{Introduction}
\begin{figure*}
\includegraphics[scale=0.5]{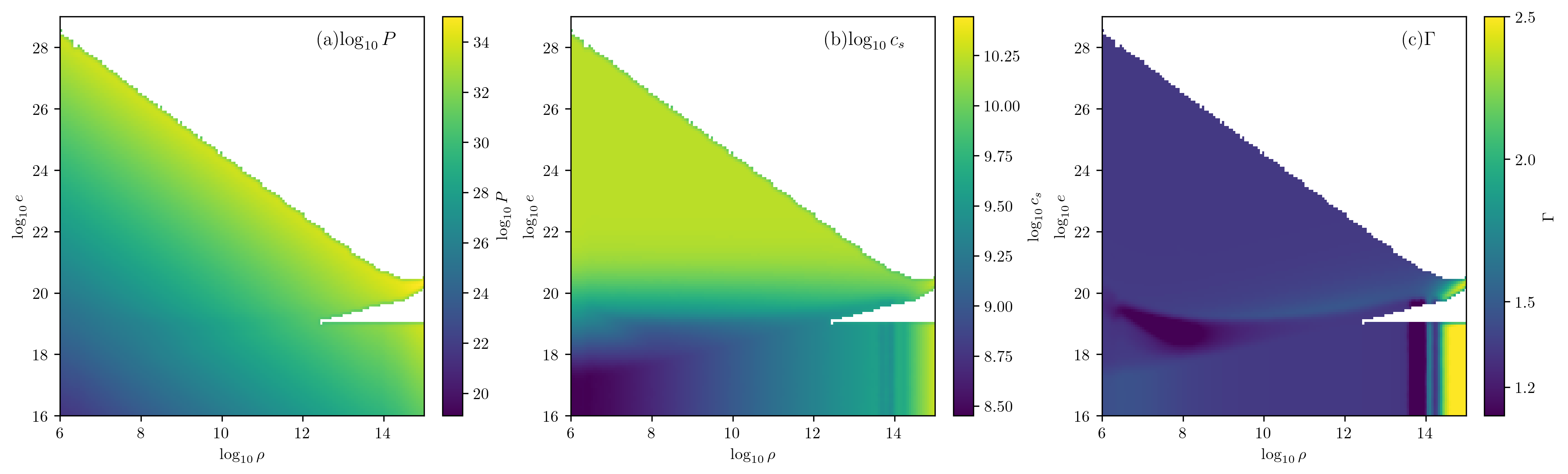}
\caption{Thermodynamic structure of the LS220 EOS at fixed electron fraction $Y_e = 0.4$. Shown are (a) $\log P$, (b) $\log c_s$, and (c) $\Gamma$ in the $(\log\rho,\ \log e)$ plane. {Alt text: Three color maps of pressure, sound speed, and effective adiabatic index. Pressure varies smoothly across the domain, while the sound speed and adiabatic index show more complex structure.}}
\label{fig:eosmap}
\end{figure*}

Nuclear equations of state (EOSs) play a central role in core-collapse supernova (CCSN) simulations and neutron-star astrophysics \citep[e.g.,][]{fischer+14,fischer+17,furusawa+23}. Modern multidimensional CCSN simulations commonly employ tabulated nuclear EOSs, such as the Shen EOS \citep{shen+98}, which provide thermodynamic quantities over the wide range of density, temperature, and electron fraction encountered during stellar collapse and the subsequent evolution \citep[see also][]{lattimer+91,hempel+10,hempel+2012,steiner+13,togashi+17,furusawa+17,shen+20,shen+25}.

The nuclear EOS affects many aspects of the CCSN dynamics, including the collapse and bounce, proto-neutron-star structure, shock evolution, neutrino emission, and gravitational-wave signals \citep[e.g.,][]{sumiyoshi+05,sumiyoshi+2023,suwa+13,yashin+20,boccioli22,nakazato+22,jakobus+23,murphy+24,rusakov+26}. Reliable nuclear EOS tables have therefore become an indispensable component of modern multidimensional CCSN simulations
\citep[e.g.,][]{kotake+12,powell+20,masada+22,matsumoto+22,matsumoto+24,kuroda+24,burrows+24,janka25}.

In such simulations, the EOS is queried throughout the hydrodynamic evolution to determine quantities such as the pressure and sound speed from the local thermodynamic state. A realistic nuclear EOS is typically stored as a multidimensional table covering a large thermodynamic domain, and the table must be retained in memory and repeatedly accessed during the calculation. The computational implications of such table lookups have motivated recent efforts to improve EOS interpolation and its performance on modern computing architectures \citep[e.g.,][]{miller+24,hammond+25}. At the same time, the memory required to store the EOS can also become a practical concern in large-scale calculations, particularly on architectures where the available memory per computational device is limited. This motivates a complementary question: {\it can the table itself be replaced by a much more compact function while retaining the accuracy required by the hydrodynamics?}

Neural networks provide a natural possibility because they can approximate complex nonlinear mappings with a relatively small number of parameters. Machine-learning surrogate models have recently been explored for a variety  of computationally intensive problems in astrophysics, including radiative transfer and reduced-order emulation \citep{kerzendorf+21,himes+22,sethuram+23,licata+23}. For a nuclear EOS, however, pointwise regression accuracy alone is not sufficient. The surrogate is evaluated repeatedly within a nonlinear hydrodynamic solver, where errors in pressure, characteristic wave speeds, and related thermodynamic quantities can directly affect the evolution  \citep{toro09,menikoff+89}. A useful EOS surrogate must therefore be assessed not only by how accurately it reproduces the original table, but also by whether it preserves the hydrodynamic evolution.

In this paper, we investigate neural-network surrogates for tabulated nuclear EOSs in CCSN hydrodynamics, using the Lattimer--Swesty EOS with incompressibility $K=220\,{\rm MeV}$ (LS220; \citealt{lattimer+91}) as a reference. The surrogate directly predicts the pressure $P$, effective adiabatic index $\Gamma$, and sound speed $c_s$ from $(\log_{10}\rho,\log_{10}e,Y_e)$, where $\rho$ is the mass density, $e$ is the specific internal energy, and $Y_e$ is the electron fraction. We assess the thermodynamic accuracy of the surrogate, including its ability to reconstruct rapid variations associated with phase transitions, and evaluate its performance in both dynamical and hydrostatic configurations. Specifically, we perform one-dimensional shock-tube tests over different thermodynamic conditions, examine how regression accuracy relates to hydrodynamic fidelity, and compare equilibrium configurations obtained by solving the Tolman--Oppenheimer--Volkoff equations. We also quantify the reduction in memory required to store the EOS and demonstrate the deployment of the surrogate in a one-dimensional adiabatic core-collapse calculation.

\section{Neural-Network EOS Surrogate}
\begin{table*}
\centering
\caption{Performance of representative NN EOS surrogate models with different loss weights, node counts, and network depths. ``Parameters'' denotes the total number of trainable parameters, and ``Val.~MSE'' is the mean squared error evaluated on the validation set. Median and 99th-percentile ($p_{99}$) relative errors are shown for $(P,\Gamma,c_s)$.\ The fiducial model used for the thermodynamic reconstruction in this work is highlighted in bold. ``Rel.~cost'' denotes the EOS evaluation cost relative to the tabulated EOS in the shock-tube implementation, while ``hydro\ $L_1(P)$'' denotes the pressure $L_1$ error measured in the hydrodynamic test described in Sec.~\ref{sec:hydro-validation}.
}
\label{tab:model_comparison}
\begin{tabular}{ccccccccc}
\hline
Weights & Nodes & Layers & Parameters & Val. MSE & Median rel. error
& $p_{99}$ rel. error & Rel. cost & hydro $L_1(P)$\\
\hline
$(1,1,1)$ & 128 & 4 &  $5.0\times10^4$ &
$9.8\times10^{-5}$ &
$(5.3,0.24,1.0)\times 10^{-3}$ &
$(4.7,1.3,1.9)\times 10^{-2}$ & 13 & $7.6\times 10^{-4}$\\

$(10,1,8)$ & 128 & 4 &  $5.0\times10^4$ &
$1.4\times10^{-4}$ &
$(2.8,0.29,0.64)\times 10^{-3}$ &
$(2.8,2.0,1.9)\times 10^{-2}$ & 13 & $1.5\times 10^{-3}$\\

\textbf{$(10,1,8)$} & \textbf{128} & \textbf{3} &
\textbf{$3.4\times10^4$} &
\textbf{$2.7\times10^{-4}$} &
\textbf{$(3.4,0.46,0.80)\times 10^{-3}$} &
\textbf{$(3.0,3.6,2.6)\times 10^{-2}$} &
\textbf{6.1} &
\textbf{$7.3\times 10^{-4}$}\\

$(10,1,8)$ & 256 & 4 &  $2.0\times10^5$ &
$7.7\times10^{-5}$ &
$(2.9,0.31,0.64)\times 10^{-3}$ &
$(2.3,1.1,1.5)\times 10^{-2}$ & 31 & $1.4\times 10^{-3}$\\

$(10,1,8)$ & 256 & 3 &  $1.3\times10^5$ &
$1.4\times10^{-4}$ &
$(2.5,0.36,0.59)\times 10^{-3}$ &
$(2.3,1.9,1.8)\times 10^{-2}$ & 69 & $6.7\times 10^{-4}$\\

$(10,1,8)$ & 64 & 4 &  $1.3\times10^4$ &
$3.3\times10^{-4}$ &
$(5.1,0.56,1.2)\times 10^{-3}$ &
$(4.8,4.0,2.9)\times 10^{-2}$ & 1.3 & $9.3\times 10^{-4}$\\
\hline
\end{tabular}
\end{table*}
Our goal is to replace the multidimensional EOS table used in hydrodynamic calculations with a compact neural network while preserving the thermodynamic information required by the fluid evolution. We use the LS220 EOS table as the reference and construct a surrogate that maps the local thermodynamic state $(\log_{10}\rho,\log_{10}e,Y_e)$ directly to the pressure $P$, effective adiabatic index $\Gamma$, and sound speed $c_s$. These quantities enter the hydrodynamic evolution through pressure forces and characteristic wave speeds and therefore provide the thermodynamic
closure required by the Godunov-type solver \citep{swesty96,timmes+00,toro09}.

We generate training samples across the full thermodynamic domain of the reference LS220 EOS table using a mixture of global random sampling (33.3\%) and targeted sampling in the low-internal-energy ($10^{14}\leq e\leq10^{19}\ {\rm erg\ g^{-1}}$: 33.3\%), high-electron-fraction ($0.35\leq Y_e\leq0.50$: 23.3\%), and overlapping low-energy/high-$Y_e$ (10.0\%) regimes. The targeted sampling provides denser coverage in regions where accurate reconstruction was found to be more challenging, while the density is sampled across the full tabulated range in all cases. For each sampled state, the original LS220 routine is used to evaluate the target quantities $(\log_{10}P,\log_{10}\Gamma,\log_{10}c_s)$, and only states for which the EOS interpolation succeeds are retained. The resulting dataset contains $5\times10^6$ valid EOS samples, spanning $\rho=3.16\times10^5$--$2.51\times10^{15},{\rm g,cm^{-3}}$, $T=0.100$--$98.75,{\rm MeV}$, and $Y_e=0.0100$--$0.5012$.

The specific internal energy requires some care when constructing this mapping. Its absolute normalization can depend on the EOS implementation and the adopted energy convention \citep{fischer+14,oertel+17}.\footnote{In the present LS220 implementation, the zero point of $e$ is defined relative to iron, following the convention adopted in the reference EOS table.} Here, $e$ is treated as the thermodynamic energy variable used by the conservative hydrodynamic equations, and the surrogate is trained using exactly the same convention as the reference LS220 implementation. The surrogate therefore reproduces the mapping actually queried by the hydrodynamic solver, without requiring an independent definition of the absolute energy zero point.

The input and output variables are standardized using the mean and standard deviation of the training set. Our fiducial surrogate is a fully connected multilayer perceptron with three hidden layers and 128 neurons per layer, using hyperbolic tangent activation functions. The network simultaneously predicts $\log_{10}P$, $\log_{10}\Gamma$, and $\log_{10}c_s$. Alternative network sizes and architectures are examined in Sec.~\ref{sec:eos_accuracy} to determine how model complexity affects both thermodynamic accuracy and hydrodynamic behavior.

The network is trained by minimizing the weighted mean-squared error
\begin{equation}
\mathcal{L} = w_P\mathcal{L}_P + w_\Gamma\mathcal{L}_\Gamma + w_{c_s}\mathcal{L}_{c_s},
\end{equation}
with $(w_P,w_\Gamma,w_{c_s})=(10,1,8)$ for the fiducial model. Pressure and sound speed are given larger weights because they directly affect pressure forces and characteristic wave speeds in the hydrodynamic solver \citep{toro09}. Together with the targeted sampling described above, the weighting is used to improve the reconstruction of these quantities while retaining coverage of the full thermodynamic domain.

The resulting network contains only $\simeq3.4\times10^4$ trainable parameters, compared with the multidimensional LS220 table that must otherwise be stored and accessed during the calculation. Whether such a compact mapping is sufficiently accurate for hydrodynamic evolution, however, cannot be established from the training loss alone. We therefore evaluate the surrogate in the following sections at two levels: first by comparing its reconstructed thermodynamic structure with the reference EOS, and then by directly testing its behavior in hydrodynamic calculations.

\begin{figure} 
\includegraphics[scale=0.15]{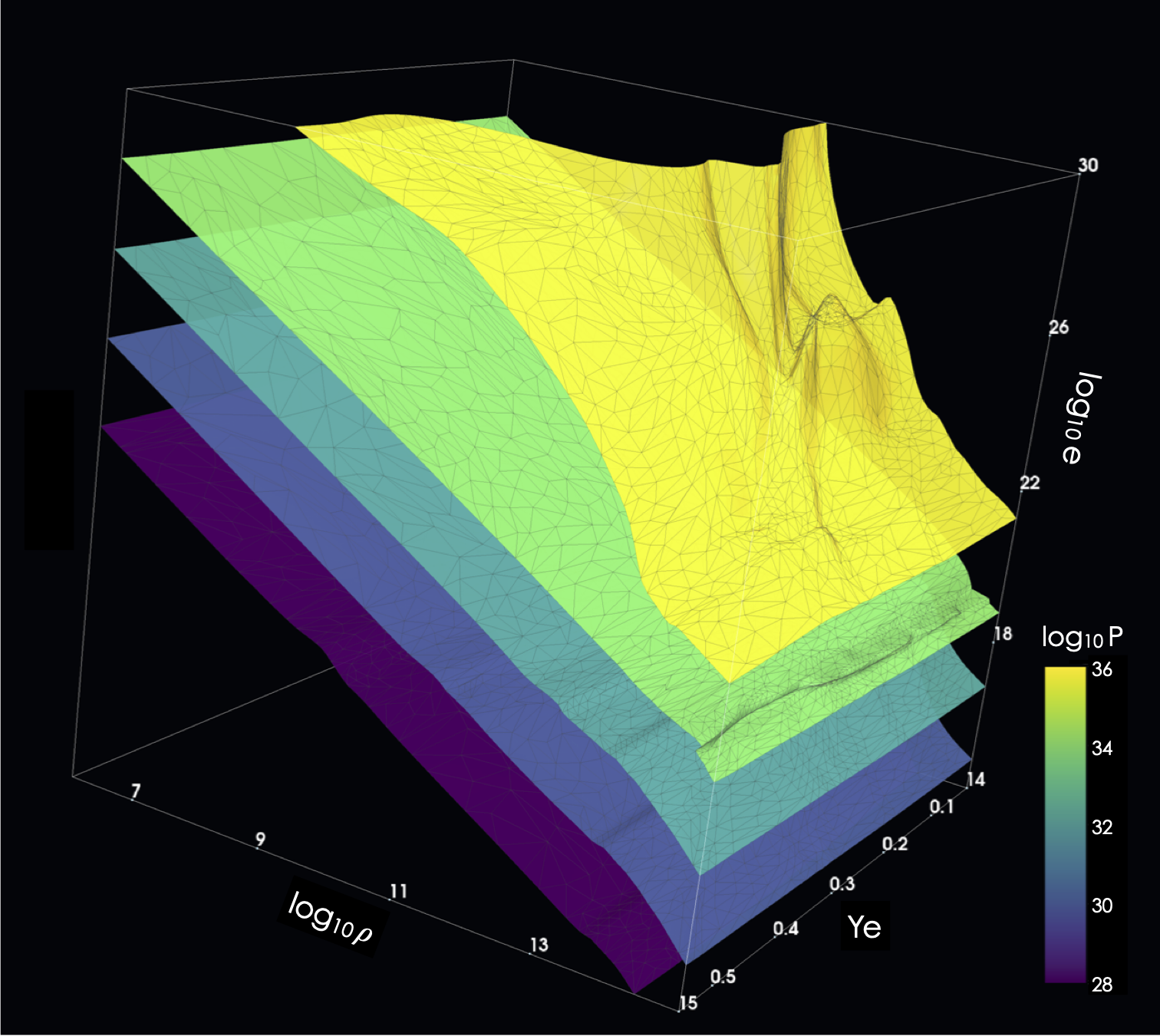} 
\caption{Global thermodynamic structure reconstructed by the fiducial neural surrogate. Isosurfaces show the NN-predicted pressure in the three-dimensional input space $(\log\rho,\log e,Y_e)$. {Alt text: Three-dimensional view of four pressure isosurfaces predicted by the neural surrogate. The isosurfaces form smooth, curved sheets across the thermodynamic domain.}} 
\label{fig:3d_manifold} 
\end{figure} 

\section{EOS Structure and Surrogate Accuracy}
\label{sec:eos_accuracy}
\subsection{Global Thermodynamic Structure}
\begin{figure*}
\includegraphics[scale=0.5]{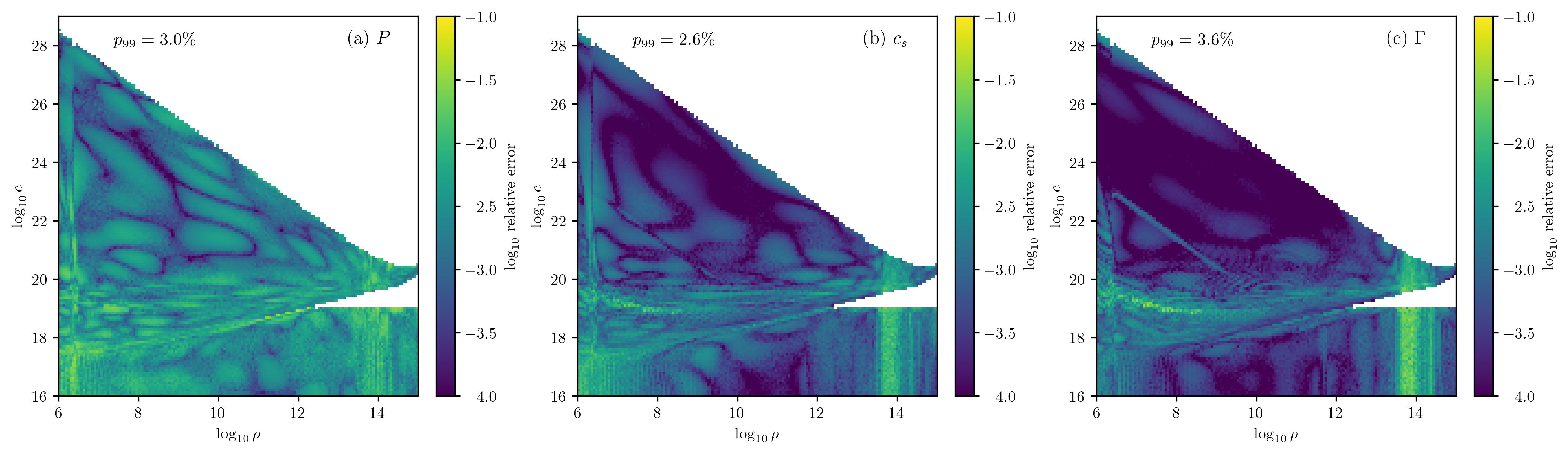} 
\caption{Relative error maps of the fiducial NN surrogate for (a) $P$, (b) $c_s$, and (c) $\Gamma$ at fixed $Y_e=0.4$. Larger residuals are confined mainly to high-density, low-energy regions where the EOS exhibits the strongest nonlinear structure. {Alt text: Three color maps of relative errors in pressure, sound speed, and effective adiabatic index. Errors are small over most of the domain, with larger errors localized mainly at high density and low internal energy.}} 
\label{fig:errormap} 
\end{figure*} 

\begin{figure} 
\includegraphics[scale=0.65]{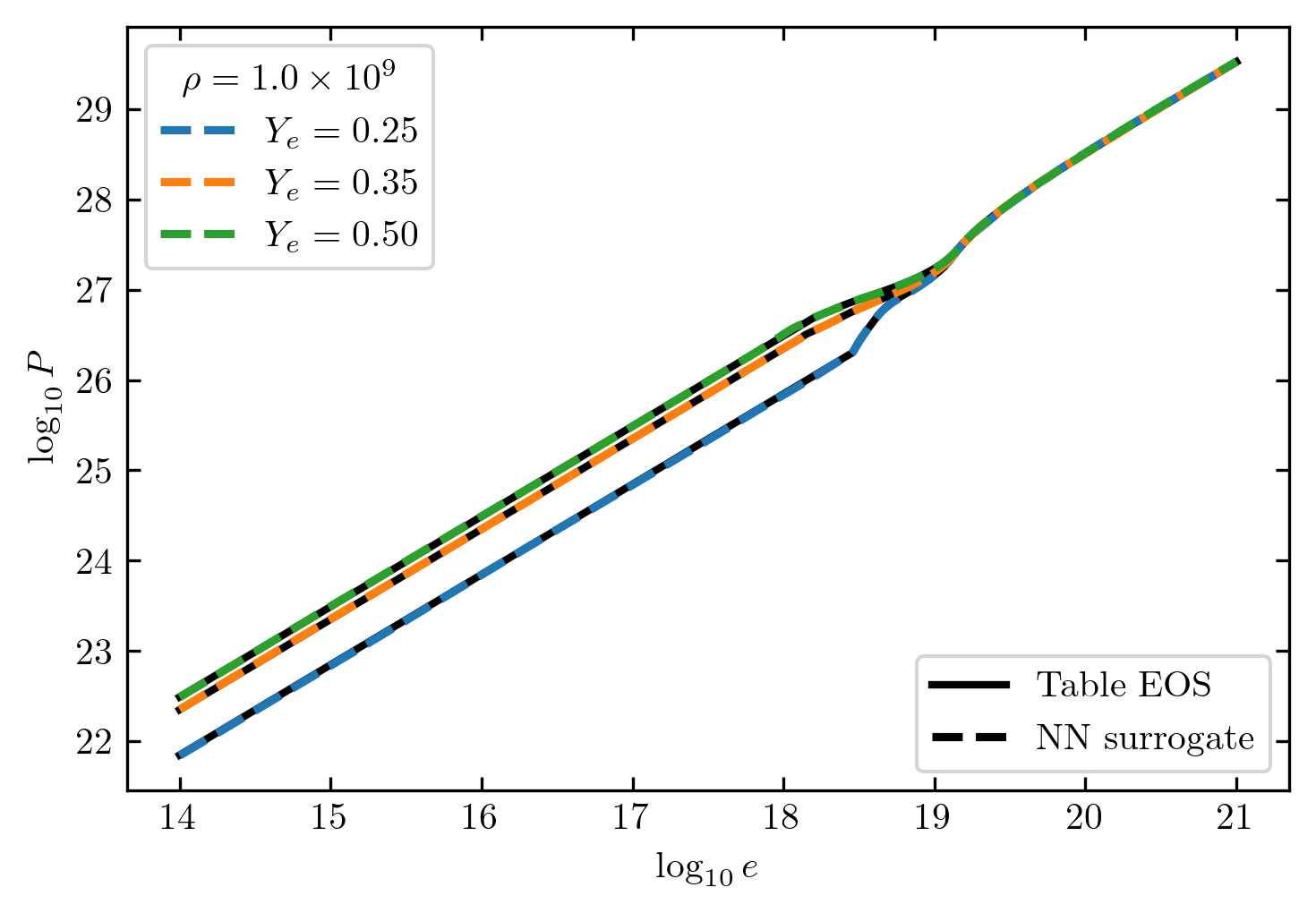} 
\caption{Pressure as a function of internal energy for representative electron fractions at fixed density ($\rho = 10^9\,{\rm g\,cm^{-3}}$). Black solid curves denote the original LS220 EOS, while colored dashed curves show the NN surrogate. The surrogate accurately reproduces the pressure--energy relation of the reference EOS, including the nonlinear transition around $\log_{10}e\sim18.5$. {Alt text: Pressure versus specific internal energy for several electron fractions. The neural-surrogate curves closely follow the LS220 curves, including their nonlinear variation.}} 
\label{fig:pressure_manifold} 
\end{figure}

Figure~\ref{fig:eosmap} illustrates the thermodynamic structure that the neural surrogate must reproduce. Representative slices of the LS220 EOS are shown at fixed electron fraction in the $(\log\rho,\log e)$ plane for pressure, sound speed, and effective adiabatic index. The pressure spans a broad dynamic range, while the sound speed develops strong nonlinear variations across the thermodynamic domain. The effective adiabatic index is comparatively smooth over much of the domain but exhibits more pronounced structure toward high density and low internal energy, where the EOS becomes strongly nonlinear \citep{lattimer+91,oertel+17}. The three target quantities therefore present different levels of difficulty for a common surrogate model, motivating the weighted training described in Sec.~2.

Table~\ref{tab:model_comparison} summarizes the reconstruction accuracy of representative networks with different loss weights and architectures. All models were trained for 5000 epochs, by which time the validation losses had converged. The validation mean squared error is evaluated using the standardized output variables, while the median and $p_{99}$ relative errors characterize the error distributions of the physical quantities over the validation set. Even the uniformly weighted model, $(w_P,w_\Gamma,w_{c_s})=(1,1,1)$, reconstructs all three thermodynamic quantities with percent-level or better accuracy over most of the sampled domain. Assigning larger weights to $P$ and $c_s$ redistributes the reconstruction accuracy among the three quantities, allowing the training to place greater emphasis on quantities that directly enter the hydrodynamic evolution.

For the detailed thermodynamic analysis below, we adopt the network with $(w_P,w_\Gamma,w_{c_s})=(10,1,8)$, 128 neurons per hidden layer, and three hidden layers as the fiducial model. It contains approximately $3.4\times10^4$ trainable parameters. As shown in Table~\ref{tab:model_comparison}, several different network architectures achieve comparable thermodynamic accuracy, and increasing the network size does not lead to a uniform improvement in all measures of performance. The relation between network size, regression accuracy, and hydrodynamic behavior will be examined further in Sec.~\ref{sec:hydro-validation}.

Figure~\ref{fig:3d_manifold} shows a global view of the thermodynamic mapping reconstructed by the fiducial surrogate. Isosurfaces of the NN-predicted pressure are displayed in the three-dimensional input space $(\log\rho,\log e,Y_e)$. The reconstructed pressure surfaces show that a network with only $\simeq3.4\times10^4$ parameters can represent the large-scale structure of the EOS across the full three-dimensional input domain, including its increasingly nonlinear behavior toward high density.

\subsection{Surrogate Accuracy and Model Dependence}
A more quantitative assessment is provided in Figure~\ref{fig:errormap}, which shows the relative errors of the fiducial surrogate at fixed $Y_e=0.4$. The 99th-percentile relative errors over the full validation set are approximately $3.0\%$, $3.6\%$, and $2.6\%$ for $P$, $\Gamma$, and $c_s$, respectively, while the corresponding median errors are all well below $1\%$. The largest residuals are concentrated mainly in the high-density, low-internal-energy region, where the reference EOS exhibits its strongest nonlinear structure. Over the majority of the thermodynamic domain, the surrogate closely reproduces the reference EOS.

Figure~\ref{fig:pressure_manifold} provides a direct comparison between the reference EOS and the neural surrogate along representative thermodynamic slices. At fixed density $\rho=10^9\,{\rm g\,cm^{-3}}$, the pressure is shown as a function of internal energy for several values of $Y_e$. The surrogate closely follows the LS220 pressure--energy relation throughout the plotted range, including the nonlinear variation around $\log_{10}e\sim18.5$. The agreement across different electron fractions indicates that the network reconstructs the continuous thermodynamic mapping between the sampled variables rather than only reproducing isolated training states.

These results establish that a relatively small neural network can reproduce the multidimensional thermodynamic structure of the reference EOS with high pointwise accuracy. For use in a CCSN hydrodynamics code, however, this is only the first requirement. The EOS is queried repeatedly along dynamically selected thermodynamic trajectories, and small local errors can influence pressure forces and wave propagation throughout the evolution. We therefore next test whether the reconstructed EOS preserves the hydrodynamic solution itself.
\begin{figure*}
\includegraphics[scale=0.45]{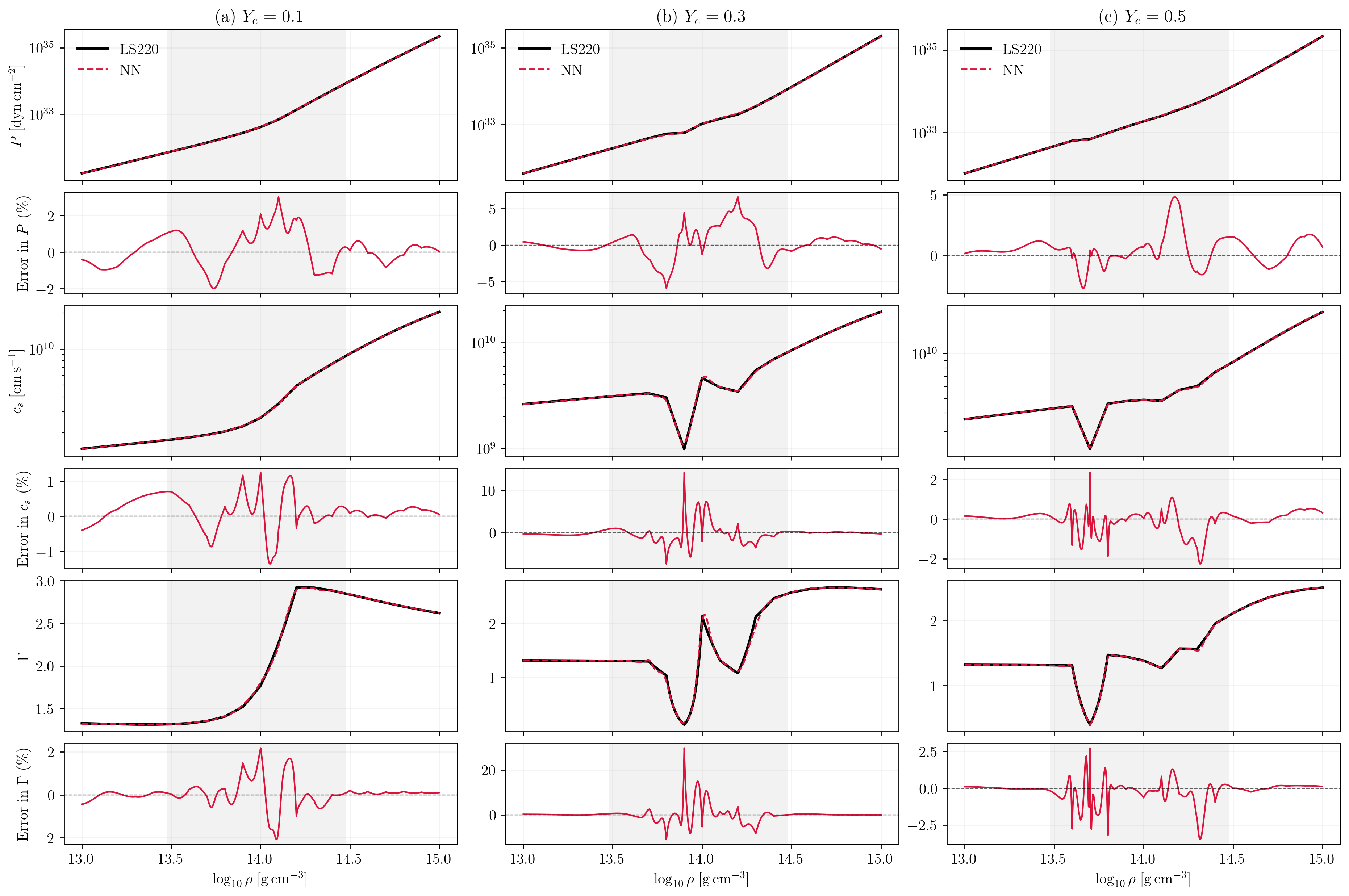}
\caption{Comparison of thermodynamic quantities along isentropic paths with $s=1,k_{\rm B}$ per baryon for (a) $Y_e=0.1$, (b) $Y_e=0.3$, and (c) $Y_e=0.5$. From top to bottom, the panels show $P$, $c_s$, and $\Gamma$, each followed by its relative error. Black solid and red dashed curves represent the LS220 EOS and neural surrogate, respectively. Relative errors are defined as $100(X_{\rm NN}/X_{\rm LS220}-1)$, where $X=P$, $c_s$, or $\Gamma$. Shaded regions indicate $3\times10^{13}\leq\rho\leq3\times10^{14},{\rm g,cm^{-3}}$. {Alt text: Eighteen-panel comparison of LS220 and neural-surrogate predictions for three electron fractions. Each column shows pressure, sound speed, and adiabatic index with their relative errors. The curves largely overlap, with localized discrepancies near sharp transitions, particularly in $\Gamma$ at $Y_e=0.3$. Shading highlights the transition region.}}
\label{fig:phase_transition}
\end{figure*}

\subsection{EOS Reconstruction across Phase Transitions}
\label{sec:phase_transition}
The nuclear EOS exhibits rapid variations in its thermodynamic properties across phase-transition regions, where the effective adiabatic index and sound speed can change substantially over a relatively narrow density interval \citep[e.g.,][]{lattimer+91, shen+98, Chamel+08,oertel+17}. Such regions provide a particularly demanding test for neural EOS surrogates, since accurate reconstruction requires the network to reproduce not only the overall thermodynamic trends but also localized nonlinear features.

To assess the surrogate accuracy in these regions, we compare the pressure $P$, sound speed $c_s$, and effective adiabatic index $\Gamma$ along isentropic thermodynamic paths obtained from the reference LS220 EOS. We adopt a fixed entropy of $s=1\,k_{\rm B}$ per baryon and consider three electron fractions, $Y_e=0.1$, $0.3$, and $0.5$, over the density range $10^{13}\leq\rho\leq10^{15}\,{\rm g\,cm^{-3}}$. For each density and electron fraction, the corresponding specific internal energy is determined from the LS220 EOS under the prescribed entropy condition. The same $(\rho,e,Y_e)$ states are then supplied to the neural surrogate, ensuring a direct comparison along identical thermodynamic trajectories.

Figure~\ref{fig:phase_transition} presents the resulting thermodynamic profiles and their relative deviations from
the reference EOS. The LS220 profiles exhibit pronounced changes in $c_s$ and $\Gamma$ over relatively narrow density intervals, reflecting the strongly nonlinear thermodynamic behavior associated with phase transitions. The neural surrogate reproduces these features, including the locations and overall amplitudes of the rapid variations, while maintaining close agreement with the reference pressure profiles.

The relative errors become more pronounced near the sharpest thermodynamic variations, particularly for $c_s$ and $\Gamma$. Among the three isentropic trajectories, the largest deviations occur at $Y_e=0.3$, where the maximum relative errors reach $6.6\%$, $14.2\%$, and $29.7\%$ for $P$, $c_s$, and $\Gamma$, respectively. The relatively large percentage error in $\Gamma$ arises near a pronounced local minimum of the reference EOS ($\Gamma\simeq0.13$), where the neural surrogate reproduces the sharp dip with a smoother and shallower profile. The corresponding absolute difference between the NN and LS220 values is $\Delta\Gamma\simeq0.04$. These discrepancies are highly localized: the relative error in $\Gamma$ exceeds $10\%$ at only 15 out of 501 sampled density points ($\sim3.0\%$) and exceeds $20\%$ at only
one point ($\sim0.2\%$). For $Y_e=0.1$ and $0.5$, the maximum relative errors remain below $5\%$ for all three quantities.

These results demonstrate that the neural surrogate reproduces the overall thermodynamic structure along the examined isentropic trajectories, while retaining localized inaccuracies near the sharpest variations. An important question is whether such local discrepancies in the EOS representation appreciably affect the global equilibrium structure. We address this issue in Sec.4.2 by comparing hydrostatic configurations obtained from the Tolman--Oppenheimer--Volkoff (TOV) equations using the reference and surrogate pressure relations.

\section{Validation of the Neural EOS Surrogate}\label{sec:hydro-validation}
\subsection{Shock-Tube Tests}\label{sec:shock-tube}
\begin{figure}
\includegraphics[scale=0.42]{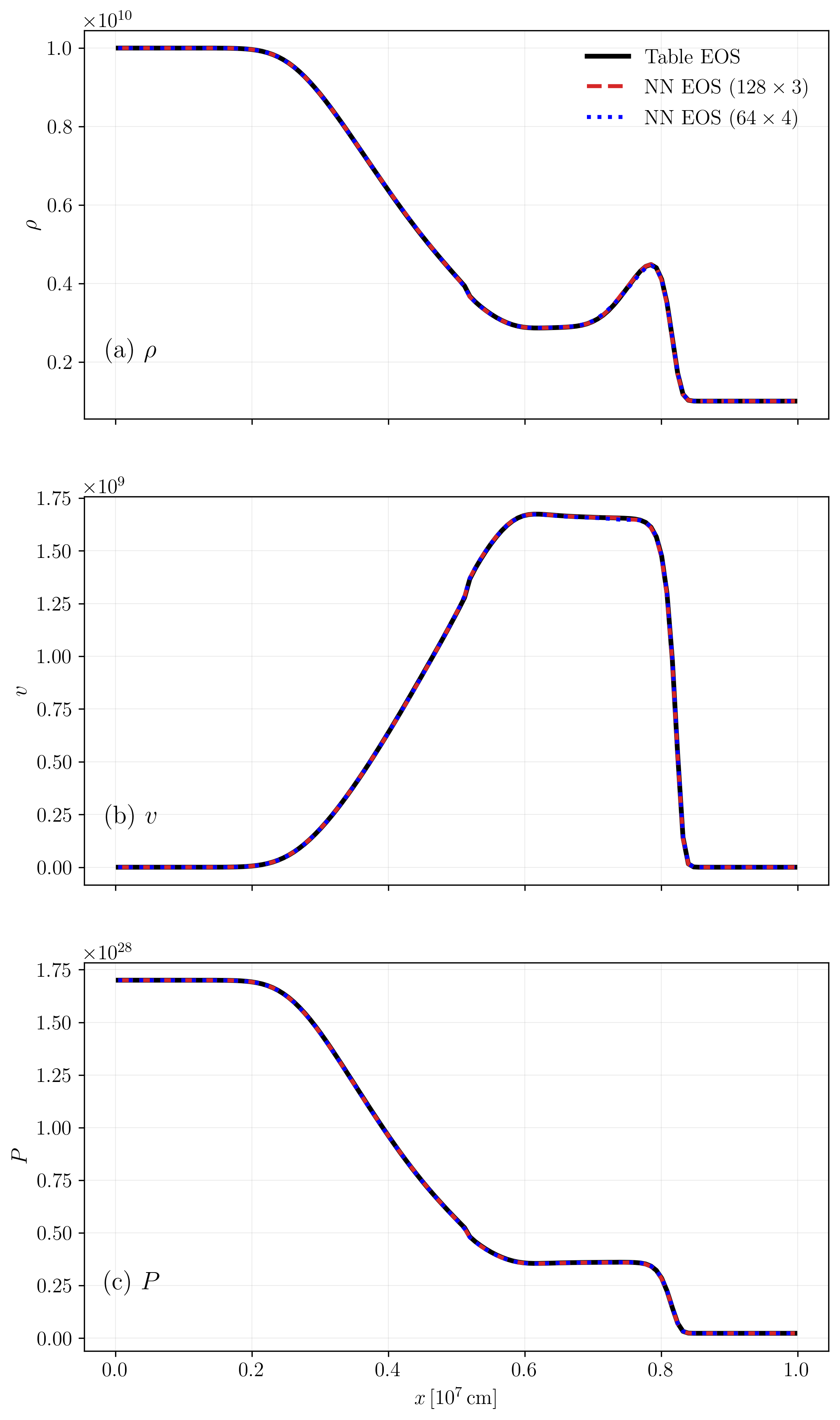}
\caption{Shock-tube solutions at $t=1.5\times10^{-3}\,\mathrm{s}$ obtained with the original LS220 table (black solid curves), the fiducial $128\times3$ neural surrogate (red dashed curves), and the compact $64\times4$ surrogate (blue dotted curves). Panels show (a) density, (b) velocity, and (c) pressure. The relative $L_1$ errors in $(\rho,v,P)$ are $(6.5\times10^{-4},\,1.4\times10^{-3},\,7.3\times10^{-4})$ for the fiducial model and $(1.0\times10^{-3},\,1.9\times10^{-3},\,9.3\times10^{-4})$ for the compact model. {Alt text: Density, velocity, and pressure profiles in the shock-tube test. The LS220 and two surrogate solutions nearly overlap throughout the domain.}}
\label{fig:shocktube}
\end{figure}
Accurate reconstruction of the EOS does not by itself guarantee that the surrogate will preserve the hydrodynamic evolution. We therefore test the neural EOS directly in a one-dimensional shock-tube problem, where errors in pressure and characteristic wave speeds can affect the propagation of shocks and other hydrodynamic structures. The surrogate is coupled directly to a Godunov-type hydrodynamic solver \citep{toro09,colella+84} and replaces the tabulated EOS during the evolution.

The calculations are performed in a one-dimensional Cartesian domain of length $L=10^7\,{\rm cm}$ with $N_x=128$ uniform cells. The initial discontinuity is placed at $x=L/2$, and zero-gradient outflow boundary conditions are imposed at both ends. The simulations are evolved to $t=1.5\times10^{-3}\,{\rm s}$ with a Courant factor of 0.35.

The left and right states are specified by
\begin{equation}
(\rho,P,Y_e)_L = (10^{10}\,{\rm g\,cm^{-3}}, 1.7\times10^{28}\,{\rm dyn\,cm^{-2}},0.40)
\end{equation}
and
\begin{equation}
(\rho,P,Y_e)_R = (10^{9}\,{\rm g\,cm^{-3}}, 2.3\times10^{26}\,{\rm dyn\,cm^{-2}},0.40),
\end{equation}
respectively, with zero initial velocity on both sides. Because the EOS surrogate is parameterized by $(\rho,e,Y_e)$, the corresponding specific internal energy is determined during initialization by solving $P(\rho,e,Y_e)=P_{\rm target}$ at fixed $(\rho,Y_e)$. The same hydrodynamic setup is then evolved using either the original LS220 table or the neural EOS, so that differences between the solutions arise from the EOS treatment.

Figure~\ref{fig:shocktube} compares the solutions at $t=1.5\times10^{-3}\,{\rm s}$. In addition to the fiducial $128\times3$ network introduced in Sec.~2, we show the $64\times4$ network as a representative compact model. Both neural surrogates closely reproduce the density, velocity, and pressure profiles obtained with the original LS220 table. Indeed, the three solutions are nearly indistinguishable on the scale of the figure, including the positions and amplitudes of the principal hydrodynamic structures.

We quantify the difference from the table-EOS solution using the relative $L_1$ error,
\begin{equation}
\epsilon_{L_1}(f) = \frac{\langle |f_{\rm NN}-f_{\rm table}|\rangle}{\langle |f_{\rm table}|\rangle},
\end{equation}
where $\langle\cdots\rangle$ denotes a spatial average over the computational domain. The relative errors remain at the $\sim10^{-3}$ level or below for all three hydrodynamic variables shown in Fig.~\ref{fig:shocktube}. For pressure, which we use for the architecture comparison in Table~\ref{tab:model_comparison}, the fiducial $128\times3$ model gives $\epsilon_{L_1}(P)=7.3\times10^{-4}$, while the much smaller $64\times4$ model gives $9.3\times10^{-4}$. Thus, reducing the network from approximately $3.4\times10^4$ to $1.3\times10^4$ trainable parameters produces only a small change in the hydrodynamic solution.

The architecture comparison in Table~\ref{tab:model_comparison} also reveals an important distinction between thermodynamic regression accuracy and hydrodynamic fidelity. For example, the $256\times4$ network achieves the smallest validation MSE, $7.7\times10^{-5}$, but its shock-tube pressure error, $1.4\times10^{-3}$, is larger than that of both the fiducial and compact models. Conversely, the $64\times4$ model has a larger validation MSE but retains pressure accuracy at the $\sim10^{-3}$ level in the hydrodynamic solution. The validation loss alone therefore does not provide a sufficient criterion for selecting an EOS surrogate for hydrodynamic calculations. Direct hydrodynamic validation is therefore important for determining whether improved pointwise reconstruction translates into improved fluid evolution.

The hydrodynamic accuracy of the compact model is particularly notable given its small memory requirement. The $64\times4$ network contains approximately $1.3\times10^4$ trainable parameters, requiring only about $100$~kB to store its weights and biases in double precision. By comparison, the LS220 table used here occupies approximately $160$~MB. The neural surrogate therefore reduces the memory required to store the EOS by a factor of approximately $1500$ while retaining $\sim10^{-3}$-level accuracy in the shock-tube solution.

The present implementation does not provide a corresponding advantage in evaluation speed. As summarized in Table~\ref{tab:model_comparison}, the runtime cost depends strongly on network architecture and implementation, and neural-EOS evaluation remains comparable to or slower than the original table lookup in the present benchmark. The principal computational advantage demonstrated here is therefore the substantial reduction in the memory required to store the EOS. Further optimization through vectorized or accelerator-oriented implementations may reduce the evaluation overhead, but such optimization is beyond the scope of the present study. 

To examine the robustness of the compact surrogate beyond the fiducial shock-tube configuration, we perform additional tests with different initial densities, pressures, and electron fractions using the $64\times4$ network. The results are summarized in Appendix~\ref{app:shock_tests}, where the surrogate is shown to reproduce the reference LS220 solutions with similarly small hydrodynamic errors across the examined conditions. The thermodynamic domain sampled by these shock-tube calculations is discussed further in Sec.~\ref{sec:thermodynamic_regimes}. We also compare the LS220 solutions with those obtained using an analytical EOS in Appendix~\ref{app:analytical_eos}, providing a quantitative measure of the differences between the nuclear EOS and a simplified theoretical prescription in the relevant parameter regime. 


\subsection{Tolman–Oppenheimer–Volkoff Models}
\label{sec:tov}
\begin{figure}
\includegraphics[scale=0.55]{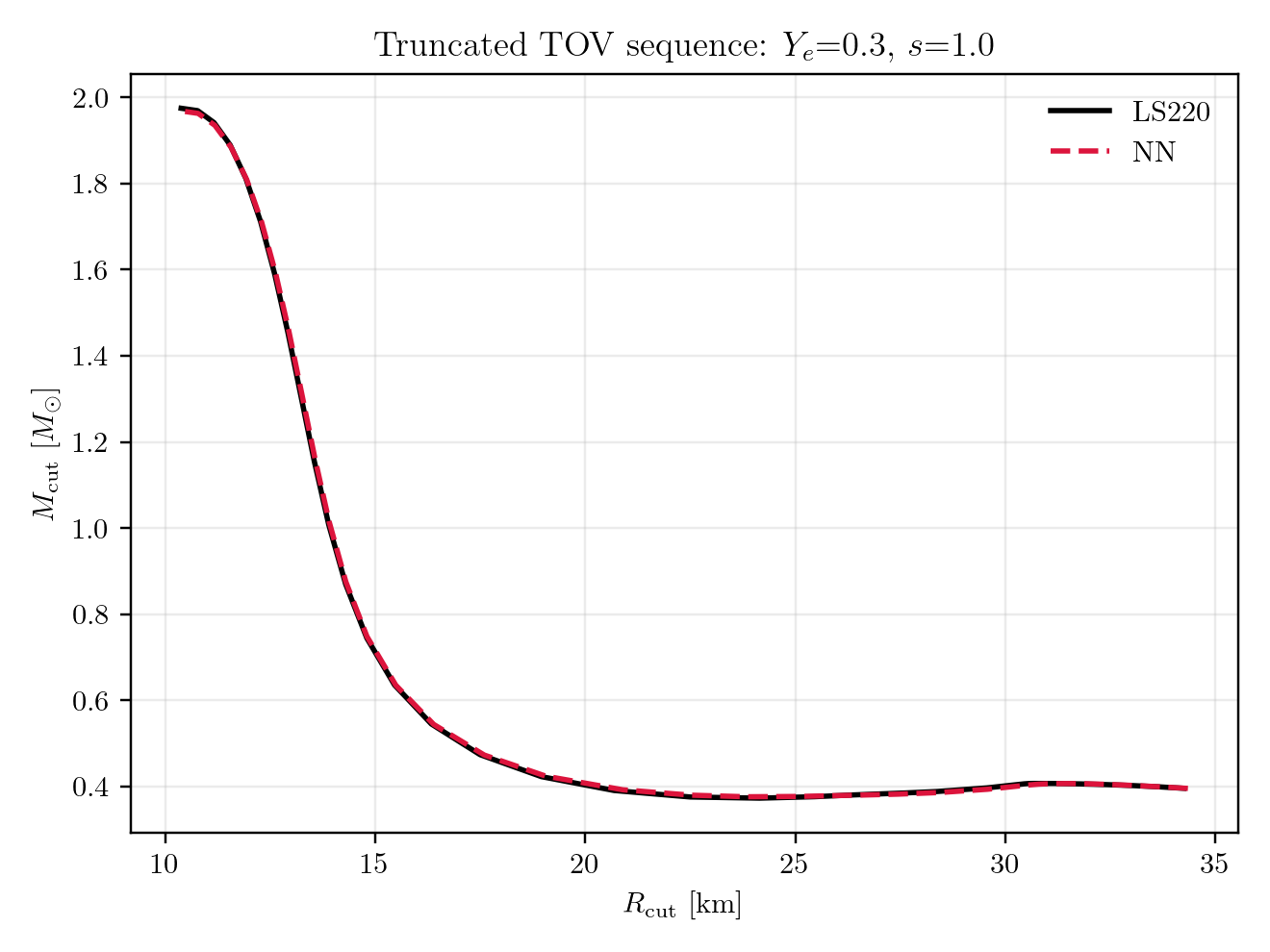}
\caption{Truncated mass--radius sequences obtained with the LS220 EOS and neural surrogate for $s=1,k_{\rm B}$ per baryon and $Y_e=0.3$. The enclosed mass $M_{\rm cut}$ and radius $R_{\rm cut}$ are evaluated at $\rho_{\rm cut}=10^{12}\ {\rm g\ cm^{-3}}$ and do not represent the total mass and radius of a neutron star. {Alt text: Truncated TOV mass--radius sequences for LS220 and the neural surrogate, showing closely overlapping curves.}}
\label{fig:TOV_M-R}
\end{figure}

\begin{figure}
\includegraphics[scale=0.48]{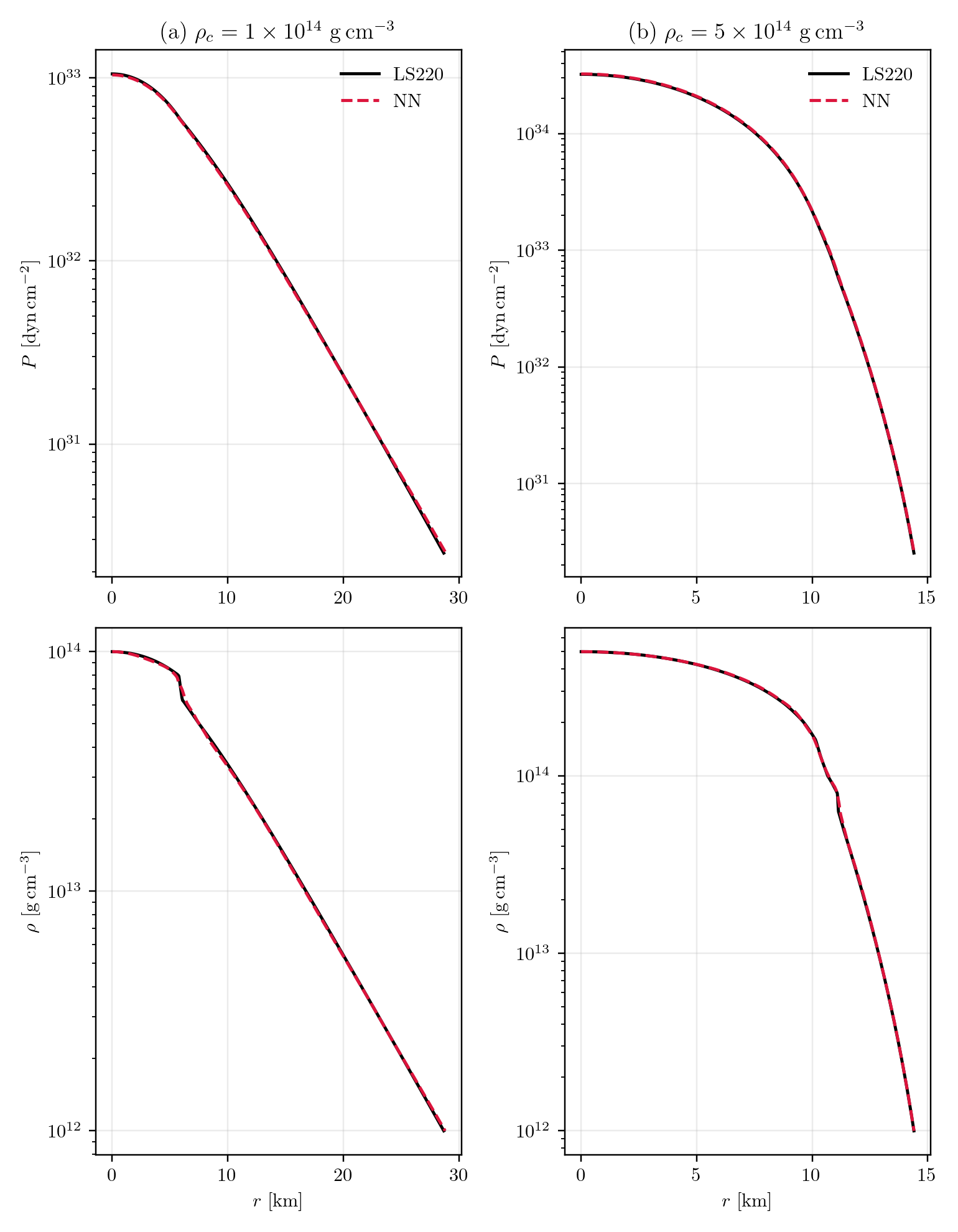}
\caption{Radial density and pressure profiles of truncated TOV models with central densities of (a) $\rho_c=10^{14}\,{\rm g\ cm^{-3}}$ and (b) $\rho_c=5\times10^{14}\,{\rm g\ cm^{-3}}$. The LS220 EOS and neural surrogate are compared for $s=1\,k_{\rm B}$ per baryon and $Y_e=0.3$, with integrations terminated at $\rho_{\rm cut}=10^{12}\,{\rm g\ cm^{-3}}$. {Alt text: Radial density and pressure profiles for two central densities, comparing LS220 and neural-surrogate TOV models. The curves largely overlap, with a localized density difference in panel (a).}}
\label{fig:TOV_profiles}
\end{figure}
As an additional test of the neural EOS surrogate, we examine its ability to reproduce the global hydrostatic structure of self-gravitating stellar cores. Unlike the shock-tube tests, which probe the dynamical response of the fluid, the Tolman--Oppenheimer--Volkoff (TOV) equations provide an integrated test of the pressure--density relation under relativistic hydrostatic equilibrium (\citealt{tolman39}; \citealt{oppenheimer+39}); see also \citealt{muller+12}. This comparison is particularly relevant because the surrogate exhibits localized discrepancies near the thermodynamic transitions discussed in Sec.~\ref{sec:phase_transition}. 

Using either the reference LS220 EOS or the neural surrogate, we construct spherical equilibrium models by integrating the TOV equations,
\begin{equation}
\frac{dP}{dr} = -\frac{G(\epsilon+P)(m+4\pi r^3P/c^2)}{c^2r^2(1-2Gm/rc^2)},
\end{equation}
and
\begin{equation}
\frac{dm}{dr} = \frac{4\pi r^2\epsilon}{c^2},
\end{equation}
where $G$ is the gravitational constant, $c$ is the speed of light, $r$ is the radial coordinate, $m(r)$ is the enclosed gravitational mass, and $\epsilon$ is the total energy density, including the rest-mass contribution.

For consistency, both calculations follow the same isentropic thermodynamic trajectory with a fixed entropy of $s=1\,k_{\rm B}$ per baryon and a uniform electron fraction of $Y_e=0.3$. At each density, the specific internal energy $e(\rho)$ is determined from the reference LS220 EOS under these prescribed conditions. The pressure is then evaluated independently using either the LS220 EOS or the neural surrogate at the same $(\rho,e,Y_e)$ state. The total energy density entering the TOV equations is taken to be $\epsilon=\rho(c^2+e)$ in both calculations. Thus, the comparison isolates the effect of the surrogate pressure representation on the resulting equilibrium structure. 

To avoid extrapolating the neural surrogate beyond the thermodynamic domain considered here, the integrations are
terminated when the density decreases to $\rho_{\rm cut}=10^{12}\ {\rm g\ cm^{-3}}$. We denote the enclosed mass and radius at this cutoff by $M_{\rm cut}$ and $R_{\rm cut}$, respectively. These quantities therefore characterize truncated equilibrium configurations rather than the total masses and radii of complete neutron stars. Both EOS representations are evaluated with identical central densities and numerical integration conditions. 

Figure~\ref{fig:TOV_M-R} compares the resulting truncated mass--radius sequences obtained with the reference and surrogate EOSs. The two sequences closely follow each other over the examined range of central densities, indicating that the neural representation of the pressure--density relation preserves the global equilibrium structure to a high degree of accuracy. This agreement is notable given the localized thermodynamic discrepancies identified in Sec.~\ref{sec:phase_transition}.

A more detailed comparison is presented in Figure~\ref{fig:TOV_profiles}, which shows the radial profiles of density and pressure for representative models with central densities of $\rho_c=10^{14}$ and $5\times10^{14}\ {\rm g\ cm^{-3}}$. The neural surrogate reproduces the overall radial structure of the reference models, with particularly close agreement in the pressure profiles. A small localized difference appears in the density profile of the lower-central-density model, near the region where the reference EOS exhibits rapid thermodynamic variations. Nevertheless, the agreement in the overall profiles and truncated mass--radius sequences indicates that these local differences in the pressure representation do not lead to substantial changes in the global equilibrium configurations examined here. 

We have also performed the same comparison for $Y_e=0.1$ and $0.5$, keeping the entropy fixed at $s=1\,k_{\rm B}$ per baryon. The corresponding results are presented in Appendix~\ref{app:tov_ye}, and show similarly good agreement between the reference and surrogate equilibrium configurations. These TOV tests provide an additional assessment of the surrogate pressure representation in relativistic hydrostatic equilibrium, complementing the dynamical shock-tube tests presented in Sec.~\ref{sec:shock-tube}.

\section{Discussion}
\label{sec:discussion}
\subsection{Implications for Neural EOS Surrogates}
\label{sec: NN surrogate implications}
The present results demonstrate that a compact neural surrogate can reproduce the thermodynamic information required by hydrodynamic evolution without retaining the full tabulated EOS in memory. More importantly, the architecture comparison reveals that pointwise thermodynamic accuracy and hydrodynamic fidelity are not necessarily optimized by the same surrogate. Networks with smaller validation errors do not always produce smaller errors in the shock-tube solution. This distinction is particularly relevant for EOS surrogates intended for core-collapse supernova simulations, where the EOS is repeatedly queried as the fluid evolves through a strongly nonlinear thermodynamic regime.

The absence of a one-to-one correspondence between validation MSE and hydrodynamic error can be understood from the different quantities measured by the two diagnostics. The validation MSE measures pointwise regression accuracy over the sampled thermodynamic domain, whereas a hydrodynamic calculation probes the EOS along dynamically selected trajectories through that domain. Errors in regions that are rarely visited by the fluid can therefore contribute to the global regression metric without appreciably affecting the hydrodynamic evolution. Conversely, relatively small but systematic errors along dynamically important trajectories may influence pressure forces and characteristic wave speeds repeatedly over many time steps.

The detailed origin of this behavior remains to be established. In addition to the magnitude and location of the EOS errors, the local structure of the reconstructed thermodynamic mapping, including its smoothness and derivatives, may affect hydrodynamic evolution \citep{menikoff+89,timmes+00,toro09}. The present calculations do not isolate these effects, and we therefore do not attribute the differences among the tested networks to any single property of the learned EOS. Nevertheless, the results demonstrate that regression metrics alone are insufficient to assess an EOS surrogate for hydrodynamic applications. Direct tests within the hydrodynamic solver provide an additional and physically relevant criterion for surrogate selection.

From the perspective of core-collapse supernova calculations, the compactness of the surrogate is also important. The $64\times4$ network retains $\sim10^{-3}$-level hydrodynamic accuracy with only $\simeq1.3\times10^4$ trainable parameters, reducing the memory required to store the EOS by a factor of approximately $1500$ relative to the LS220 table used here. Such a reduction may become useful in large-scale multidimensional calculations, particularly when the available memory per computational device is limited. The present implementation does not provide a corresponding advantage in evaluation speed, however, and the practical balance between memory use and EOS evaluation cost will depend on the target architecture and implementation. Thus, the present results primarily demonstrate the potential of neural surrogates as a compact alternative to tabulated EOSs rather than as a means of accelerating EOS evaluation.

Although LS220 is used here as a reference EOS, the same strategy is in principle applicable to other tabulated nuclear EOSs. The achievable compression and hydrodynamic accuracy will depend on the complexity of the underlying thermodynamic manifold and should be assessed separately for each EOS. Extension to additional thermodynamic quantities required by more complete neutrino-radiation hydrodynamics will also be necessary for realistic core-collapse supernova applications.

\subsection{Application to Adiabatic Core-Collapse Simulations}
\label{sec:1d_core_collapse}
As a first step beyond the controlled shock-tube problem, we also performed a one-dimensional adiabatic core-collapse calculation in which the neural EOS surrogate is used throughout the hydrodynamic evolution. The calculation is performed with the {\small 3DnSNe} hydrodynamics code \citep{takiwaki+16}, using the $20\,M_\odot$ presupernova model of \citet{woosley+07} as the initial condition. The radial domain is discretized with $512$ logarithmically spaced grid cells. The electron fraction $Y_e$ is advected with the fluid, without source terms from electron capture or other weak interactions. The collapse is triggered by the loss of pressure support associated with the photodisintegration of iron. Gravity is treated using the effective relativistic potential of \citet{marek2006}, adopting the Case A prescription to approximately account for general-relativistic effects.

\begin{figure}
\includegraphics[scale=0.38]{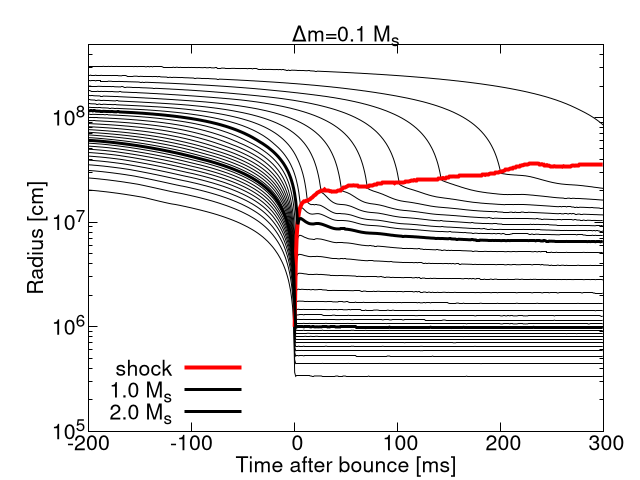}
\caption{One-dimensional adiabatic core-collapse calculation performed with the neural EOS surrogate. Black curves show mass-shell trajectories spaced by $\Delta m=0.1\,M_\odot$, with the $1\,M_\odot$ and $2\,M_\odot$ shells highlighted by thicker curves; the red curve shows the shock position. Time is measured relative to core bounce. The calculation remains stable through collapse, bounce, shock formation, and at least 300~ms of post-bounce evolution. {Alt text: Radius--time diagram showing mass-shell trajectories and the shock in the core-collapse calculation. The mass shells contract toward bounce, after which the shock propagates outward.}}
\label{fig:adiabatic_collapse}
\end{figure}

Figure~\ref{fig:adiabatic_collapse} shows representative mass-shell trajectories and the shock position as functions of time relative to core bounce. The neural EOS remains numerically stable through collapse, core bounce, shock formation, and subsequent post-bounce evolution for at least 300~ms. This calculation subjects the surrogate to a substantially broader and more strongly evolving thermodynamic trajectory than the shock-tube test and demonstrates that it can be deployed throughout the major hydrodynamic stages of a core-collapse calculation.

A corresponding calculation with the original LS220 table has not yet been performed with otherwise identical numerical settings. We therefore do not interpret Fig.~\ref{fig:adiabatic_collapse} as a quantitative validation of the collapse or post-bounce dynamics. Rather, it provides a first deployment test of the surrogate in a core-collapse calculation. Quantitative comparisons against the tabulated EOS, inclusion of weak interactions and neutrino transport, and multidimensional core-collapse simulations constitute important next steps toward assessing neural EOS surrogates under realistic supernova conditions.\\

\subsection{Thermodynamic Regimes Explored by the Hydrodynamic Simulations}
\label{sec:thermodynamic_regimes}
\begin{figure}
\includegraphics[scale=0.45]{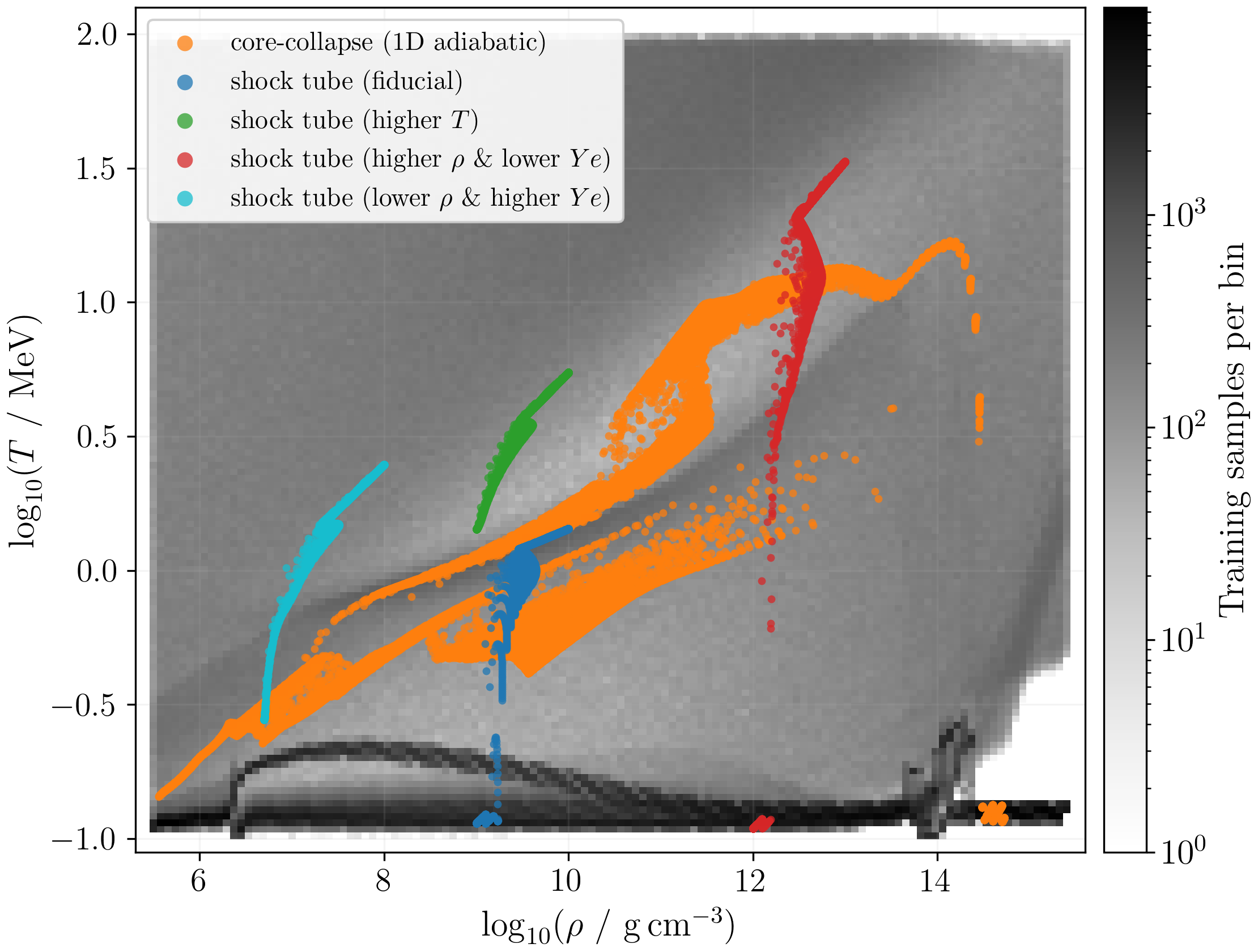}
\caption{Thermodynamic states explored by the hydrodynamic simulations in the $\rho$--$T$ plane. The grayscale background indicates the number of training samples per bin on a logarithmic scale. Colored points show the states sampled during the one-dimensional adiabatic core-collapse calculation (orange) and the four shock-tube tests (blue, green, red, and cyan). {Alt text: Density--temperature diagram showing EOS training sample density in grayscale and thermodynamic states from core-collapse and shock-tube simulations in color.}}
\label{fig:rho_T_state_space}
\end{figure}
The thermodynamic conditions encountered during hydrodynamic simulations provide an important perspective on the applicability of the neural EOS surrogate. Although the network is trained over a broad thermodynamic domain, individual hydrodynamic tests sample only particular regions of this domain. It is therefore useful to examine how the thermodynamic trajectories explored by the simulations relate to the distribution of the training data. 

Figure~\ref{fig:rho_T_state_space} shows the thermodynamic states sampled during the shock-tube tests and the one-dimensional adiabatic core-collapse calculation in the $\rho$--$T$ plane. The grayscale background represents the distribution of the EOS training samples, while the colored points indicate the states encountered during the hydrodynamic calculations. The fiducial shock-tube test (blue) probes a relatively narrow region of the thermodynamic state space. The additional shock-tube tests presented in Appendix~\ref{app:shock_tests} extend the coverage toward higher temperatures (green), higher densities with lower electron fractions (red), and lower densities with higher electron fractions (cyan).

In contrast, the one-dimensional core-collapse calculation explores a substantially broader thermodynamic regime, extending from the relatively low-density conditions of the presupernova core to the high-density and high-temperature states reached during collapse and after core bounce. The resulting trajectories span several orders of magnitude in density and a wide range of temperatures, demonstrating that the surrogate can be evaluated throughout the strongly evolving thermodynamic conditions encountered in this calculation. 

It is important, however, to distinguish thermodynamic coverage from quantitative hydrodynamic validation. The shock-tube calculations provide direct comparisons between the neural surrogate and the reference LS220 EOS within the regions sampled by those tests. The core-collapse calculation, on the other hand, demonstrates numerical stability over a much broader thermodynamic regime, but does not yet establish quantitative agreement with an otherwise identical simulation using the tabulated EOS. Consequently, the present results support the feasibility of applying neural EOS surrogates to core-collapse hydrodynamics, while further quantitative validation across the relevant thermodynamic domain remains necessary.

The initial conditions and temperature ranges of the shock-tube tests are summarized in Table~\ref{tab:shock_tube_tests} of Appendix~\ref{app:shock_tests}. In the present implementation, the neural surrogate is evaluated directly without an explicit check of whether the input state lies within the training domain. No special treatment, such as extrapolation control or fallback to the original EOS table, is applied to states outside this domain. Consequently, predictions outside the training domain rely on extrapolation, and their accuracy is not guaranteed. Explicit domain checks and fallback procedures will be important for future applications.

\section{Summary}
We have investigated neural-network surrogates for tabulated nuclear equations of state in core-collapse supernova hydrodynamics, using the LS220 EOS as a reference. The surrogate reconstructs pressure, sound speed, and effective adiabatic index from $(\log\rho,\log e,Y_e)$ and accurately reproduces the multidimensional thermodynamic structure of the reference EOS, including the rapid variations associated with phase transitions, although localized discrepancies remain. When coupled directly to a hydrodynamic solver, the surrogate reproduces one-dimensional shock-tube solutions with relative errors at the $\sim10^{-3}$ level over the different thermodynamic conditions examined. Additional tests using the Tolman--Oppenheimer--Volkoff
equations demonstrate close agreement in the global hydrostatic equilibrium structure between the reference and surrogate EOSs. 

A key result is that smaller pointwise regression errors do not necessarily translate into smaller hydrodynamic errors. Among the network architectures examined here, models with lower validation MSE do not systematically yield
more accurate shock-tube solutions. This demonstrates that regression accuracy alone is insufficient for selecting an EOS surrogate and that direct hydrodynamic validation provides an important additional criterion.

We further find that the hydrodynamic accuracy can be retained with a compact $64\times4$ network containing only $\simeq1.3\times10^4$ trainable parameters. The resulting network requires approximately $100$~kB to store its weights and biases in double precision, compared with approximately $160$~MB for the LS220 table used here, reducing the memory required to store the EOS by a factor of approximately $1500$. The present implementation does not provide a corresponding advantage in EOS evaluation speed, indicating that its principal computational benefit is the substantial reduction in memory requirement.

Finally, the neural EOS was deployed in a one-dimensional adiabatic core-collapse calculation and remained numerically stable through collapse, core bounce, shock formation, and at least $300$ ms of post-bounce evolution. Although this calculation does not yet constitute a quantitative validation against an otherwise identical table-EOS simulation, it demonstrates that the surrogate can be used throughout the strongly nonlinear thermodynamic evolution encountered in a core-collapse calculation. These results demonstrate the potential of compact neural EOS surrogates for reducing the memory requirements of nuclear EOSs in large-scale core-collapse supernova simulations.

\begin{ack}
The authors are grateful to the anonymous referee for their insightful and constructive comments, which significantly improved the quality of the manuscript. Numerical computations were carried out on Cray XD2000 at Center for Computational Astrophysics~(CfCA), National Astronomical Observatory of Japan~(NAOJ).
\end{ack}

\section*{Funding}
This work was supported by the Fukuoka University Grant No.GR2302, and also by JSPS KAKENHI Grant Numbers JP21K03612, JP21H01088, JP23K22494, JP23K25895, JP23K03400, JP24K00631 and JP25K07374. This research was also supported by MEXT as “Program for Promoting researches on the Supercomputer Fugaku” (Structure and Evolution of the Universe Unraveled by Fusion of Simulation and AI; Grant Number JPMXP1020230406) and JICFuS.

\section*{Data availability} 
The data underlying this article will be shared on reasonable request to the corresponding author.

\bibliographystyle{plainnat}
\bibliography{references}

\appendix 
\section{Additional Shock-Tube Tests}
\label{app:shock_tests}
\begin{table*}
\caption{Initial conditions, temperature ranges, and relative $L_1$ errors for the four shock-tube tests using the compact $64\times4$ neural surrogate. The subscripts $L$ and $R$ denote the initial left and right states, respectively. Densities and pressures are given in $\mathrm{g\,cm^{-3}}$ and $\mathrm{dyn\,cm^{-2}}$, respectively. The temperature ranges are given in $\mathrm{MeV}$ and evaluated over all computational cells and time steps. The relative $L_1$ errors are measured against the reference LS220 solutions.}
\centering
\label{tab:shock_tube_tests}
\small
\setlength{\tabcolsep}{3pt}

\begin{tabular}{lcccccccccc}
\hline\hline
Case
& $\rho_L$ & $\rho_R$ & $P_L$ & $P_R$ & $Y_e$ & $T$ & $\epsilon_{L1}(\rho)$ & $\epsilon_{L1}(v)$ & $\epsilon_{L1}(P)$ & $\epsilon_{L1}(c_s)$ \\
\hline
Fiducial & $10^{10}$ & $10^{9}$ & $1.7\times10^{28}$ & $2.3\times10^{26}$ & 0.40 & $0.11$--$1.43$ & $1.0\times10^{-3}$ & $1.9\times10^{-3}$ & $9.3\times10^{-4}$ & $5.7\times10^{-3}$ \\
Higher-$T$ & $10^{10}$ & $10^{9}$ & $1.7\times10^{29}$ & $2.3\times10^{27}$ & 0.40 & $1.42$--$5.47$ & $1.6\times10^{-3}$ & $4.1\times10^{-3}$ & $2.2\times10^{-3}$ & $2.7\times10^{-3}$ \\
Higher-$\rho$, lower-$Y_e$ & $10^{13}$ & $10^{12}$ & $5.0\times10^{32}$ & $1.0\times10^{29}$ & 0.30 & $0.11$--$33.52$ & $1.1\times10^{-3}$ & $2.9\times10^{-3}$ & $1.2\times10^{-3}$ & $1.6\times10^{-3}$ \\
Lower-$\rho$, higher-$Y_e$ & $10^{8}$ & $5.0\times10^{6}$ & $5.0\times10^{27}$ & $1.0\times10^{24}$ & 0.45 & $0.28$--$2.48$ & $6.2\times10^{-4}$ & $3.5\times10^{-3}$ & $1.2\times10^{-3}$ & $5.3\times10^{-3}$ \\
\hline
\end{tabular}

\end{table*}
\begin{figure*}
\includegraphics[scale=0.24]{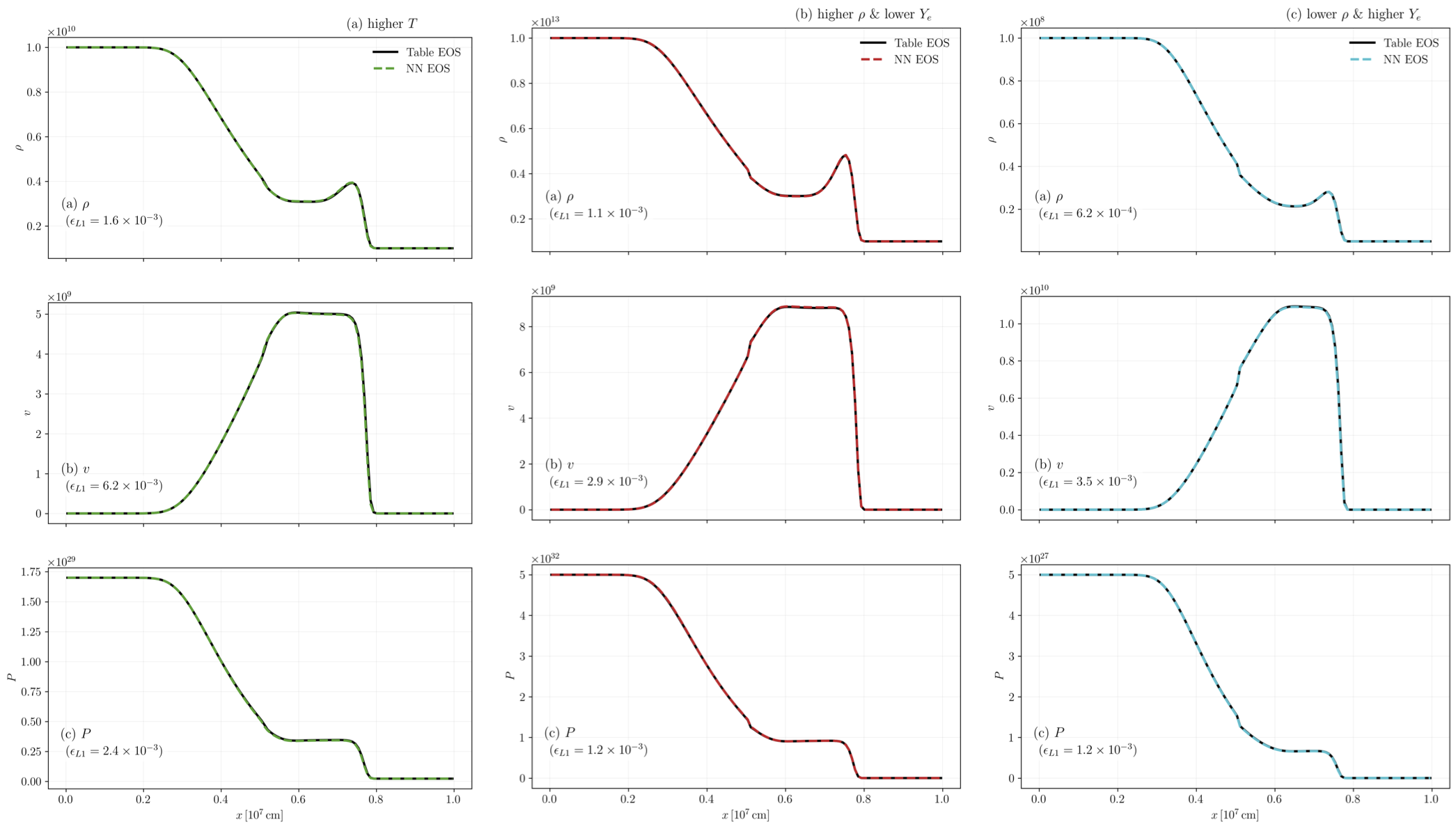}
\caption{Shock-tube solutions for three additional thermodynamic configurations: (a) higher $T$, (b) higher $\rho$ and lower
$Y_e$, and (c) lower $\rho$ and higher $Y_e$. From top to bottom, the panels show $\rho$, $v$, and $P$. Black solid and colored dashed curves represent the LS220 table and the compact $64\times4$ neural surrogate, respectively. Relative $L_1$ errors are indicated in each panel. {Alt text: Nine-panel comparison of LS220 and neural-surrogate shock-tube solutions for three thermodynamic configurations. The density, velocity, and pressure profiles nearly overlap.}}
\label{fig:shock_tube_different_states}
\end{figure*}

To examine the robustness of the neural EOS surrogate under different thermodynamic conditions, we perform three additional shock-tube tests using the compact $64\times4$ network introduced in Sec.~\ref{sec:shock-tube}. These tests extend the fiducial configuration toward higher temperatures, higher densities with lower electron fractions, and lower densities with higher electron fractions. The numerical setup is identical
to that described in Sec.~\ref{sec:shock-tube}, except for the initial thermodynamic states.

Table~\ref{tab:shock_tube_tests} summarizes the initial conditions, the temperature ranges encountered during the simulations, and the relative $L_1$ errors measured against the reference LS220 solutions. The fiducial case is also included for comparison. All calculations use the compact $64\times4$ network, and the errors are evaluated using the definition given in Sec.~\ref{sec:shock-tube}.

Figure~\ref{fig:shock_tube_different_states} compares the profiles of $\rho$, $v$ and $P$ obtained with the reference LS220 table and the neural surrogate for the three additional configurations. The two solutions closely agree in all cases, reproducing the positions and amplitudes of the principal hydrodynamic structures.

The relative $L_1$ errors in $\rho$ and $P$ remain below $2.5\times10^{-3}$ in all three additional tests, while those in $v$ remain below $4.5\times10^{-3}$. The sound-speed errors are also small, remaining below $6\times10^{-3}$ across all four configurations. These results indicate that the compact neural surrogate retains good hydrodynamic accuracy over the different thermodynamic conditions examined here.

Although these tests extend the range of conditions beyond the fiducial shock-tube configuration, they sample only selected regions of the full thermodynamic state space. Their coverage relative to the EOS training data and the one-dimensional core-collapse calculation is discussed in Sec.~\ref{sec:thermodynamic_regimes}.

\section{Comparison with Analytical EOS}
\label{app:analytical_eos}
\begin{figure}
\includegraphics[scale=0.45]{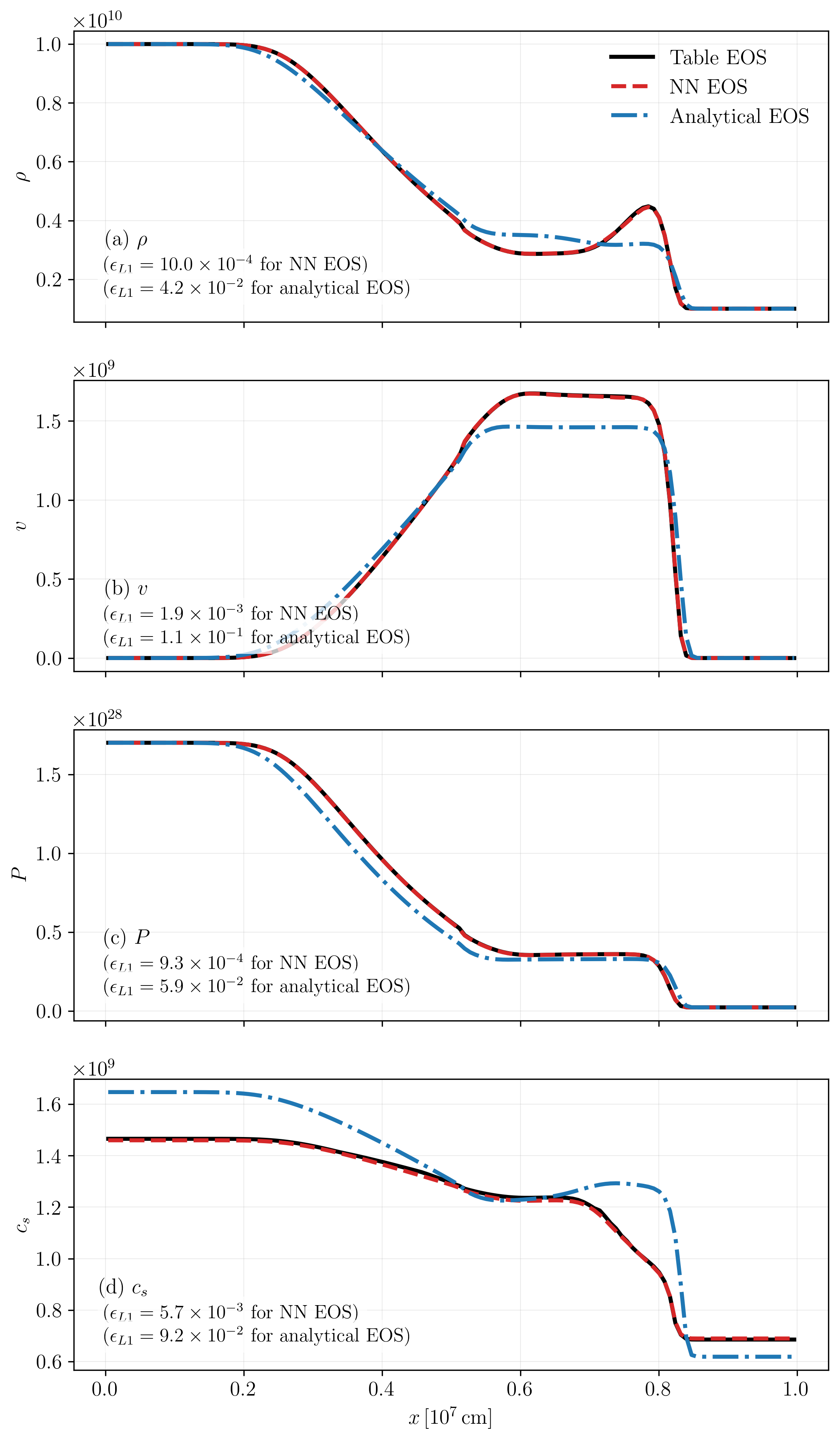}
\caption{Shock-tube solutions obtained with the LS220 table (black solid), neural surrogate (red dashed), and analytical EOS (blue dash-dotted). Panels (a)--(d) show $\rho$, $v$, $P$, and $c_s$, respectively, at $t=1.5\times10^{-3}\,{\rm s}$. {Alt text: Comparison of density, velocity, pressure, and sound-speed profiles for three EOS prescriptions.}
}
\label{fig:LS_NN_AN}
\end{figure}

\begin{figure}
\includegraphics[scale=0.45]{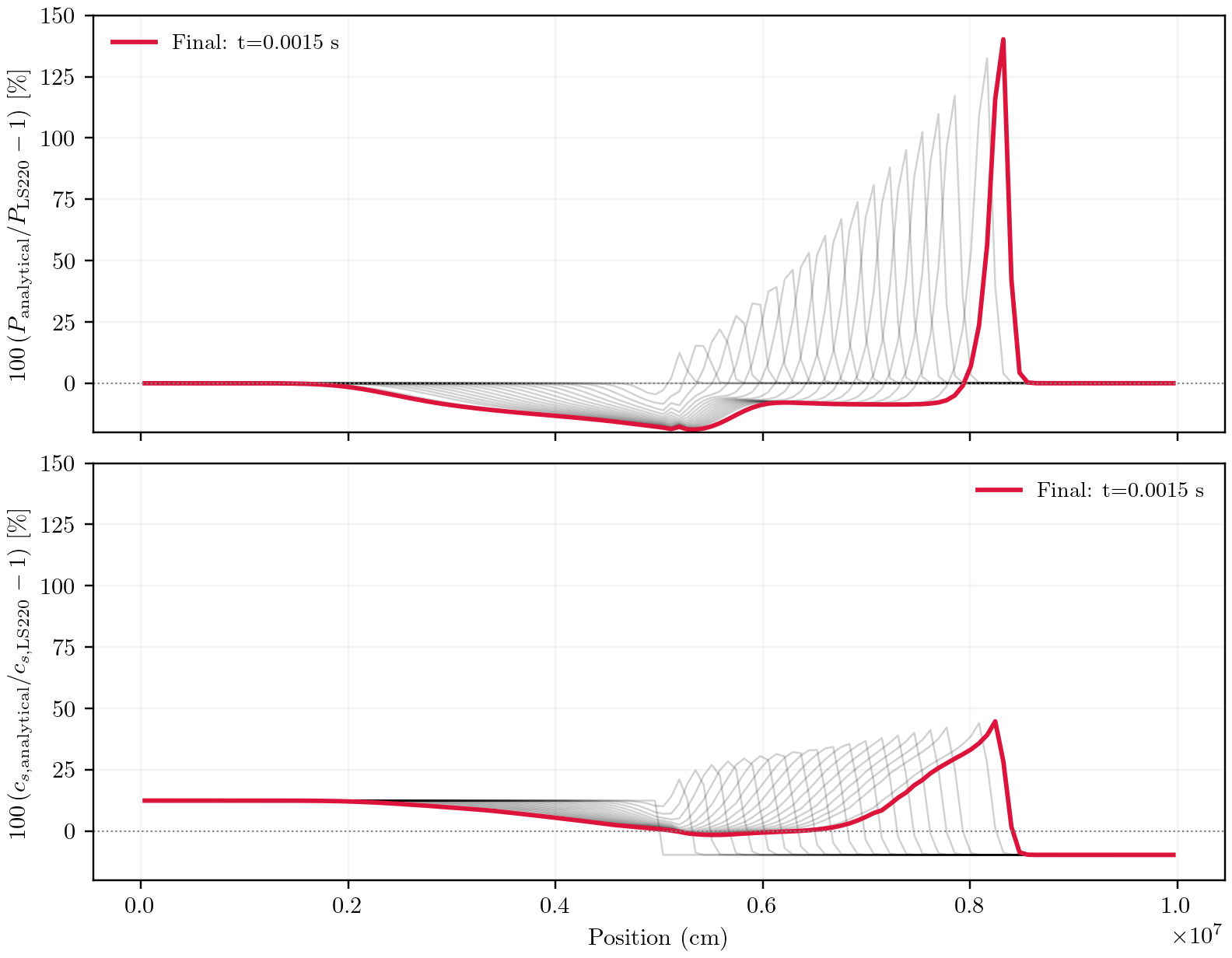}
\caption{Spatial profiles of the relative differences in
$P$ (upper) and $c_s$ (lower) between the analytical EOS and LS220 during the shock-tube evolution. Gray curves show successive times, and red curves show the final time ($t=1.5\times10^{-3}\,{\rm s}$). {Alt text: Evolution of spatial pressure and sound-speed differences between the analytical EOS and LS220.}}
\label{fig:error_LS-AN}
\end{figure}
To examine how a simplified analytical EOS differs from the reference LS220 EOS in hydrodynamic calculations, we compare shock-tube solutions obtained with the analytical EOS, LS220 table, and neural surrogate (with compact $64\times 4$ network). Following eq. (3) of \citet{yamasaki+05}, we adopt
\begin{equation}
P = \frac{11\pi^2 k_{\rm B}^4}{180c^3\hbar^3}T^4
    + \frac{\rho k_{\rm B}T}{m_{\rm N}},
\end{equation}
and
\begin{equation}
e = \frac{11\pi^2 k_{\rm B}^4}
         {60c^3\hbar^3}\frac{T^4}{\rho}
    + \frac{3k_{\rm B}T}{2m_{\rm N}},
\end{equation}
where $T$ is the temperature, $k_{\rm B}$ is the Boltzmann constant, $\hbar$ is the reduced Planck constant, $c$ is the speed of light, and $m_{\rm N}$
is the nucleon mass. The first terms represent the contributions from photons and relativistic electron--positron pairs, while the second terms describe nonrelativistic nucleons treated as an ideal gas.

We consider the fiducial shock-tube configuration described in Sec.~\ref{sec:shock-tube}, using the same numerical resolution, boundary conditions, and final time. The compact $64\times4$ neural surrogate is adopted for this comparison. The original LS220 table is used as the reference
solution, and the relative $L_1$ errors are evaluated using the definition given in Sec.~\ref{sec:shock-tube}.

Figure~\ref{fig:LS_NN_AN} compares the profiles of $\rho$, $v$, $P$ and $c_s$ at $t=1.5\times10^{-3}\,{\rm s}$. While the neural surrogate closely reproduces the reference LS220 solution, the analytical EOS produces appreciable differences in the hydrodynamic structure, including the amplitudes of the density and velocity variations and the pressure and sound-speed profiles.

The relative $L_1$ errors for the analytical EOS are $4.2\times10^{-2}$, $1.1\times10^{-1}$, $5.9\times10^{-2}$, and $9.2\times10^{-2}$ for $\rho$, $v$, $P$, and $c_s$, respectively. By comparison, the corresponding errors for the neural surrogate are $1.0\times10^{-3}$, $1.9\times10^{-3}$, $9.3\times10^{-4}$, and $5.7\times10^{-3}$. Thus, the neural surrogate yields substantially smaller deviations from the LS220 solution than the analytical EOS in all four quantities.

Figure~\ref{fig:error_LS-AN} further shows the spatial distribution of the relative differences in pressure and sound speed between the analytical EOS and LS220 at successive times. The pressure difference develops a pronounced localized peak near the propagating shock, reaching approximately $140\%$ at the final time. The sound-speed difference also becomes substantial, reaching approximately $45\%$. These large localized differences arise primarily because the shock front in the analytical-EOS solution propagates slightly ahead of those in the LS220 and neural-surrogate solutions. Even a small displacement of the shock front can produce substantial relative differences when the profiles are compared at the same spatial positions.

This comparison illustrates that replacing LS220 with a simplified analytical EOS can lead to appreciable differences in the hydrodynamic evolution, whereas the neural surrogate closely reproduces the reference solution. These findings are specific to the fiducial shock-tube configuration and the analytical prescription considered here.
\begin{figure*}
\includegraphics[scale=0.5]{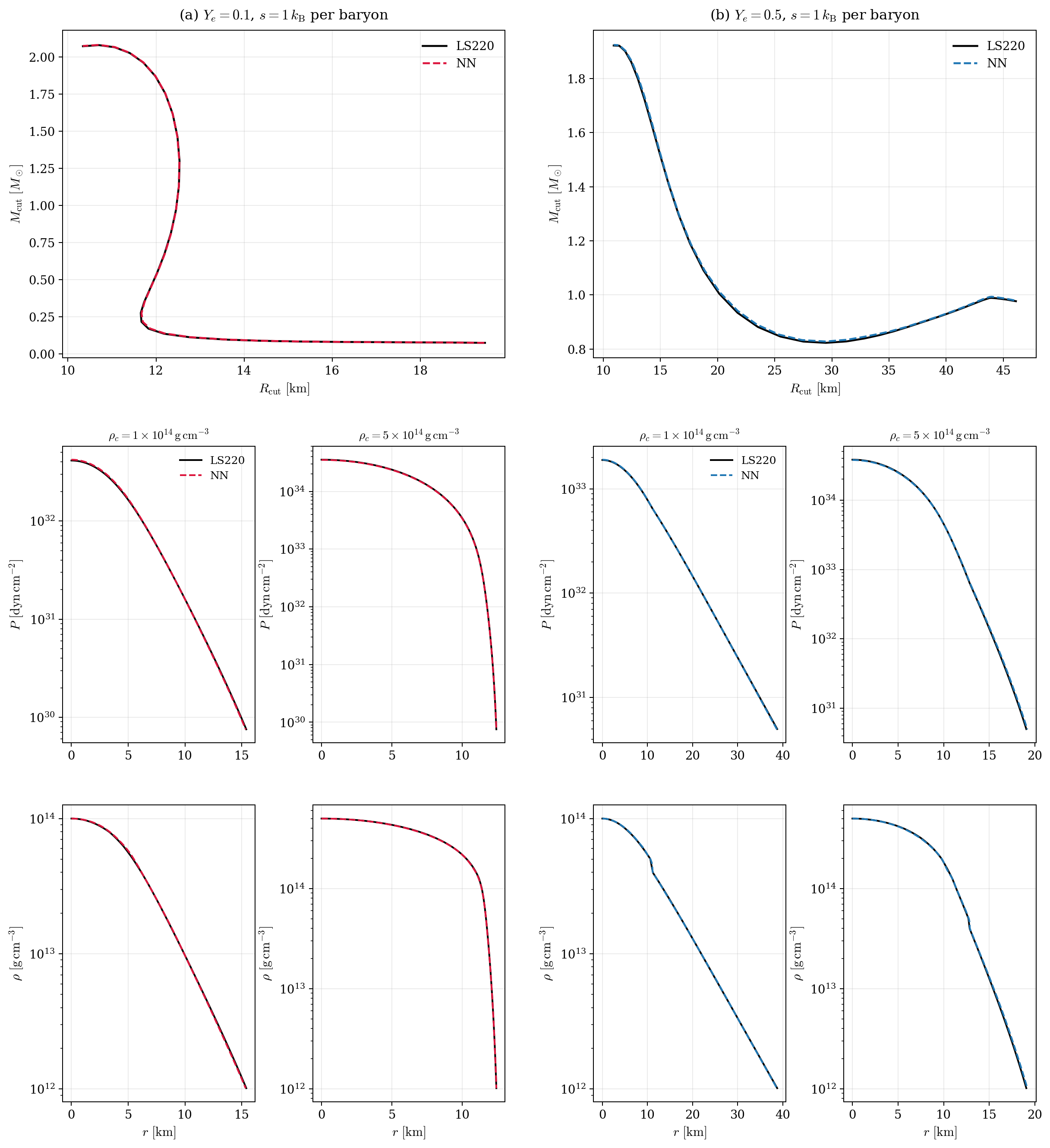}
\caption{Comparison of truncated TOV models obtained with the LS220 EOS (black solid curves) and neural surrogate (dashed curves) for (a) $Y_e=0.1$ (red) and (b) $Y_e=0.5$ (blue), with $s=1\,k_{\rm B}$ per baryon. The top panels show the truncated mass--radius sequences, while the lower panels show radial pressure and density profiles for $\rho_c=10^{14}$ and $5\times10^{14}\ {\rm g\ cm^{-3}}$. The integrations terminate at $\rho_{\rm cut}=10^{12}\ {\rm g\ cm^{-3}}$. {Alt text: Comparison of truncated TOV mass--radius sequences and radial profiles for electron fractions 0.1 and 0.5. Black LS220 and colored neural-surrogate curves nearly overlap, with small differences in the mass--radius sequences.}}
\label{fig:TOV_different_Ye}
\end{figure*}

\section{TOV solutions for different electron fraction}
\label{app:tov_ye}
To examine the dependence of the TOV results on the electron fraction, we perform additional equilibrium calculations for $Y_e=0.1$ and $0.5$, complementing the $Y_e=0.3$ models presented in Sec.~\ref{sec:tov}. The entropy is fixed at $s=1\,k_{\rm B}$ per baryon, and the numerical procedure and density cutoff are identical to those adopted in Sec.~\ref{sec:tov}. Both EOS representations follow the same LS220-derived isentropic trajectories, isolating the effect of the surrogate pressure representation.

Figure~\ref{fig:TOV_different_Ye} compares the truncated mass--radius sequences (top panels) and representative radial pressure and density profiles (lower panels) for $Y_e=0.1$ (left) and $Y_e=0.5$ (right). For $Y_e=0.1$, the neural surrogate closely reproduces the reference LS220 mass--radius sequence over the entire range examined, including the turning point at the high-mass end. Such a turning-point structure is also seen in the mass--radius relations of hot neutron stars calculated with the LS EOS by \citet{muller+12}, although their models adopt different thermodynamic conditions and are not truncated at a prescribed density. The radial profiles also show excellent agreement for both central densities, $\rho_c=10^{14}$ and $5\times10^{14}\ {\rm g\ cm^{-3}}$. For $Y_e=0.5$, small differences are visible in the mass--radius sequence, while the pressure and density profiles remain nearly indistinguishable at the scale of the figure.

Together with the $Y_e=0.3$ results in Sec.~\ref{sec:tov}, these comparisons demonstrate that the neural surrogate closely reproduces the truncated hydrostatic equilibrium configurations over the three electron fractions examined. Localized thermodynamic reconstruction errors do not produce substantial deviations in the TOV models considered here.


\end{document}